\documentclass[12pt]{article}
\usepackage{graphicx}

\newcommand\pubnumber{NuPhys2017-Giaz}
\newcommand\pubdate{\today}

\def\napoli{Department of Physics\\
Universit\'a degli studi di Padova and INFN sezione di Padova, 35131 Padova, ITALY}
\def\support{\footnote{ on behalf of JUNO collaboration
          }}

\def\Title#1{\begin{center} {\Large #1 } \end{center}}
\def\Author#1{\begin{center}{ \sc #1} \end{center}}
\def\Address#1{\begin{center}{ \it #1} \end{center}}

\newcommand\pubblock{\rightline{\begin{tabular}{l} \pubnumber\\
         \pubdate  \end{tabular}}}
\newenvironment{Abstract}{\begin{quotation}  }{\end{quotation}}
\newenvironment{Presented}{\begin{quotation} \begin{center} 
             PRESENTED AT\end{center}\bigskip 
      \begin{center}\begin{large}}{\end{large}\end{center} \end{quotation}}

\def\beq{\begin{equation}}
\def\eeq#1{\label{#1}\end{equation}}
\def\eeqn{\end{equation}}

\def\beqa{\begin{eqnarray}}
\def\eeqa#1{\label{#1}\end{eqnarray}}
\def\eeqan{\end{eqnarray}}

\let\bar=\overbar

\def\Dslash{\not{\hbox{\kern-4pt $D$}}}
\def\dslash{\not{\hbox{\kern-2pt $\del$}}}

\def\msb{{\bar{\ssstyle M \kern -1pt S}}}

\begin{document}
\begin{titlepage}
\pubblock

\vfill
\Title{Status and perspectives of the JUNO experiment}
\vfill
\Author{ Agnese Giaz \support}
\Address{\napoli}
\vfill
\begin{Abstract}
The determination of the neutrino mass hierarchy, whether the $\nu _3$ neutrino mass eigenstate is heavier or lighter than the $\nu _1$ and $\nu _2$ mass eigenstates, is one of the remaining undetermined fundamental aspects of the Standard Model in the lepton sector. Furthermore the mass hierarchy determination will have an impact in the quest of the neutrino nature (Dirac or Majorana mass terms) towards the formulation of a theory of flavour. The Jiangmen Underground Neutrino Observatory (JUNO) is a reactor neutrino experiment under construction at Kaiping, Jiangmen in Southern China composed by a large liquid scintillator detector (sphere of 35.4 m of diameter) surronding by 18000 large PMTs and 25000 small PMTs, a water cherenkov detector and a top tracker detector. The large active mass (20 kton) and the unprecedented energy resolution (3\% at 1 MeV) will allow to determine the neutrino mass hierarchy with good sensitivity and to precisely measure the neutrino mixing parameters, $\theta _{12}$, $\Delta m^2_{21} $, and $\Delta m^2_{ee}$ below the 1\% level. Moreover, a large liquid scintillator detector will allow to explore physics beyond mass hierarchy determination, in particular on many oyher topics such as in astroparticle physics, like supernova burst and diffuse supernova neutrinos, solar neutrinos, atmospheric neutrinos, geo-neutrinos, nucleon decay, indirect dark matter searches and a number of additional exotic searches. In this work the status and the perspectives of the JUNO experiment will be described, focusing also on the main physics aims and the other possible physics cases.
\end{Abstract}
\vfill
\begin{Presented}
NuPhys2017, Prospects in Neutrino Physics\\
Barbican Centre, London, UK,  December 20--22, 2017
\end{Presented}
\vfill
\end{titlepage}
\def\thefootnote{\fnsymbol{footnote}}
\setcounter{footnote}{0}

\section{Introduction on neutrino mass hierarchy measurements}

Precise measurements of the $\theta _{13}$ neutrino oscillation parameter by the Daya Bay \cite{DayaBay}, RENO \cite{Reno} and Double Chooz \cite{tre} experiments, opened the path to the determination of the neutrino mass hierarchy. Mass hierarchy determination will have an impact in the quest of the neutrino nature (Dirac or Majorana mass terms) towards the formulation of a theory of flavour. \par
The future of neutrino oscillation studies foresees a rich and wide experimental program to determinate the elements of the Pontecorvo – Maki – Nakagawa – Sakata (PNMS) oscillation matrix with unprecedented accuracy, and to get more information in the features of neutrino properties. Therefore, the Dirac or Majorana nature of the neutrino mass term, mass hierarchy determination, octant of $\theta _{23}$, violating $\delta _{cp}$ phase and improved precision of the mass-mixing parameters (mixing angles as well as squared mass differences), are the core of the oscillation measurements. \par

\begin{figure}[htb]
\centering
\includegraphics[height=2.2in]{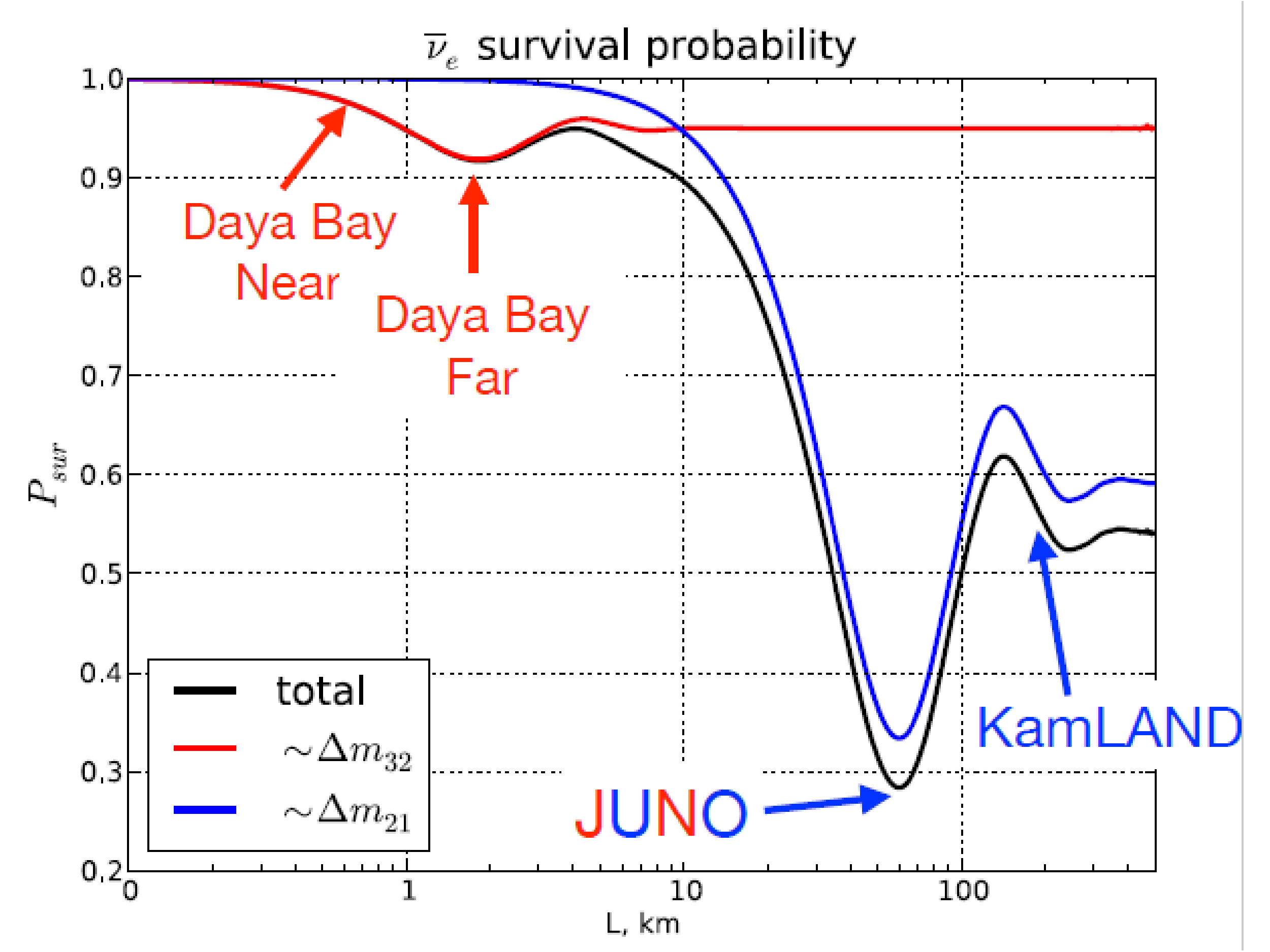}
\caption{The $\bar{\nu} _e$ survival probability as a function of the distance fron the nuclear power plant (NNP). The distance of other neutrino reactor experiments (KamLAND and Daya Bay) from the NNP are indicated and comparend with the JUNO distance at 53 km.}
\label{fig:sopr}
\end{figure}

One fundamental aspect in the lepton sector of the standard model is if the $\nu _3$ neutrino mass eigenstate is heavier or lighter than the $\nu _1$ and $\nu _2$ (mass hierarchy determination). It is known that $m_2 > m_1$ and $ |\Delta m^2_{31}| \gg |\Delta m^2_{21}|$ where $ \Delta m^2_{ij} = m^2_i - m^2_j$, but It is not known if $ m_3 > m_{1,2}$ or $ m_3 < m_{1,2}$. \par
In this context, the JUNO experiment \cite{Adam, An} will play a central role on two aspects: the determination of mass hierarchy and the precise measurements of the oscillation parameters ($sin^2 \theta _{12}$, $ \Delta m^2_{21} $, $\Delta m^2_{ee}$). \par
\begin{figure}[htb]
\centering
\includegraphics[height=2.2in]{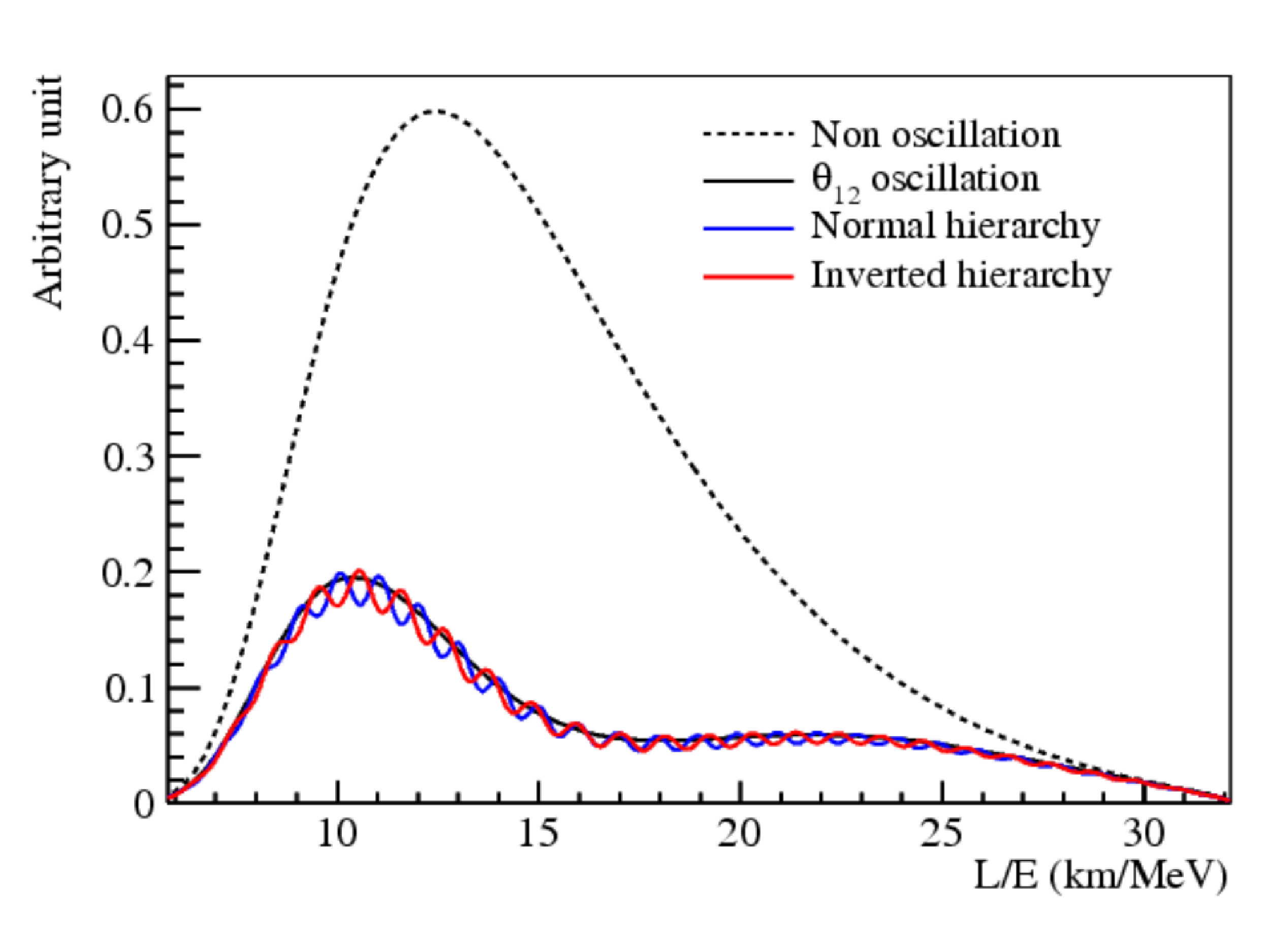}
\caption{The black line shows the un-oscillated spectra of reactor anti-neutrino at the distance of 53 km (distance betweeen JUNO site and nuclear power plants), while the red and the blue line show the oscillated spectra for the two mass orderings (blue is the normal order and red is the inverted order) \cite{An}.}
\label{fig:po}
\end{figure}

The electron antineutrino survival probability, shown in Fig. \ref{fig:sopr}, depends on the mass hierarchy even though the effect is small, but fortunately not negligible. The effect of the neutrino mass hierarchy on the electron antineutrino energy spectrum, measured at a medium-baseline of $\sim$ 53 km, is shown in Fig. \ref{fig:po} in which the normal order is represented with a blue line and the inverted one with a red line.

\section{The Jiangmen Underground Neutrino Observatory (JUNO) experiment}

The JUNO (Jiangmen Underground Neutrino Observatory) experiment is a reactor neutrino detector under construction in the south of China. The aim of the experiment is the determination of the neutrino mass hierarchy and the measurements of some neutrino oscillation parameters with a precision better than 1\%. \par
The JUNO design main goals are:
\begin{itemize}
\item Large target mass provided by 20 kton of liquid scintillator.
\item Excellent energy resolution 3\% at 1 MeV due to large light yield (1200 p.e. per MeV) and large optical coverage around 78\%;
\item Low energy scale uncertainty less than 1\%.
\end{itemize}
	
All these parameters will be improved with respect to the current neutrino experiment such as KamLAND, Borexino, Daya Bay. \par
\begin{figure}[htb]
\centering
\includegraphics[height=1.8in]{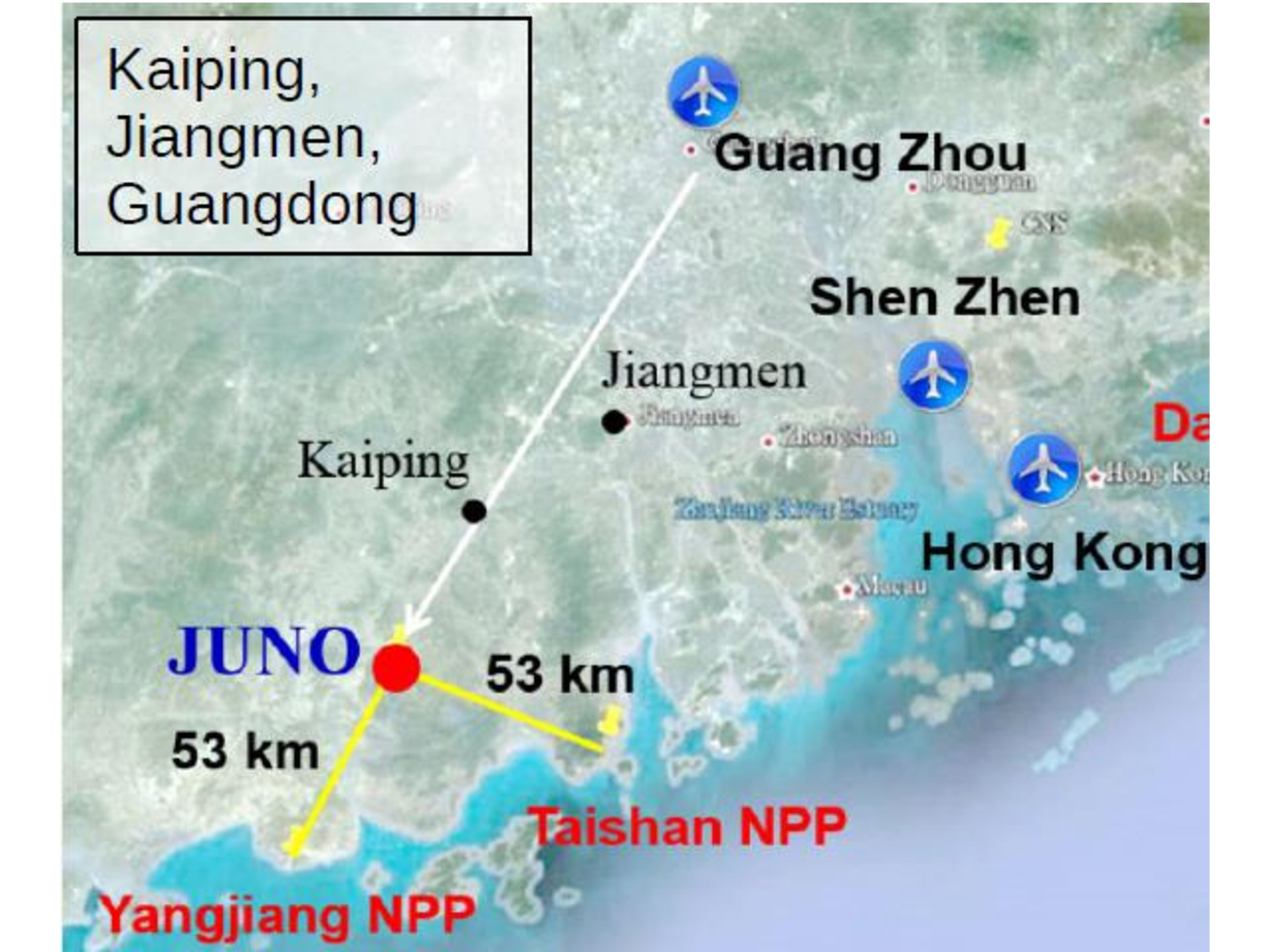}
\caption{The JUNO detector site (indicated by a red point), is located in the south of China. The two nuclear power plants at $\sim$ 53 km from JUNO experiment site are shown in the picture.}
\label{fig:site}
\end{figure}

The detector site was chosen with respect to optimal mean distance (~53 km) from the cores of 2 nuclear power plants (Yangjiang 4 cores, and Taishan 6 cores) as shown in Fig. \ref{fig:site}. The site was chosen to maximize the sensitivity to the mass hierarchy as shown in the neutrino survival probability in Fig \ref{fig:sopr}. The reactors produce a total thermal power of 35.8 GW with all reactors in operation mode and a thermal power of 26.6 GW by the start of data taking around 2020. The detector will be deployed in an underground laboratory under the Dashi hill overburden by 700 m of rocks. The experimental hall is designed to have two accesses: one is a 616 m-deep vertical shaft, and the other is a 1340 m long tunnel with a slope of 42.5\%. \par
The $\bar{\nu}_e$ flux will be detected through the reaction $\bar{\nu}_e + p \rightarrow e^+ +n $ (inverse beta decay) and its rate will be around 83 events per day.

\subsection{The JUNO detector}

\begin{figure}[htb]
\centering
\includegraphics[height=2.8in]{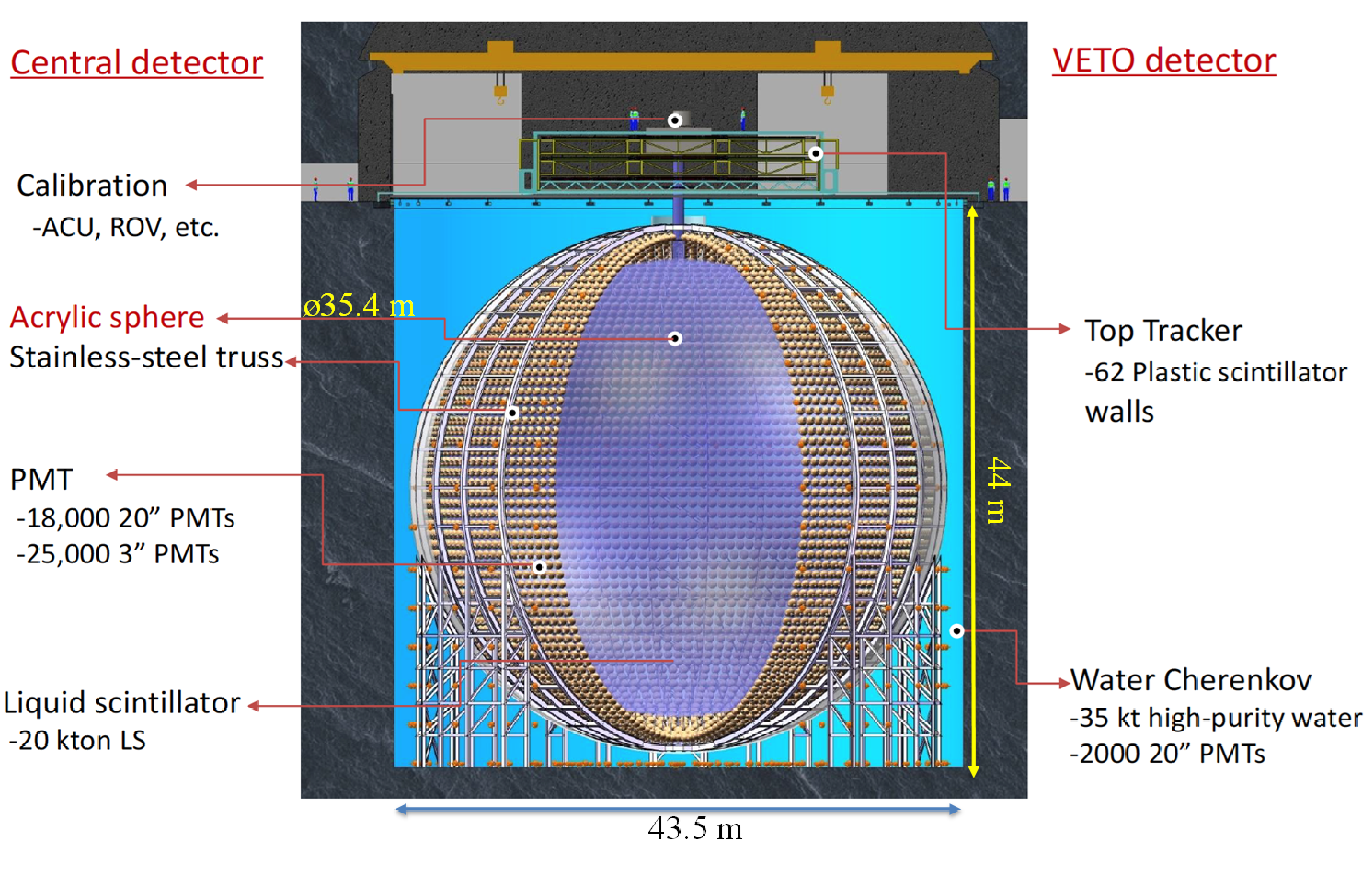}
\caption{A schematic view of the JUNO detector. The main parts are indicated: the central detector composed by the liquid scintillator, the arcrylic sphere, the PMTs and the calibration system; the veto detector composed by the water Cherenkov pool and the top tracker.}
\label{fig:render}
\end{figure}

The central detector is composed by $\sim$ 20 kton of active mass of Linear Alkyl Benzene (LAB) scintillator in an acrylic sphere of 35.4 m in diameter as shown in Fig. \ref{fig:render}. The liquid scintillator has similar recipe as the Daya Bay, it will be doped with 3 g/L 2.5-diphenyloxazole (PPO) as the fluor and 15 mg/L p-bis-(omethylstyryl)-benzene (bis-MSB) as the wavelength shifter. The detector has unprecedented energy resolution of 3\% at 1 MeV due to the light yield (1200 p.e. per MeV) and the optical coverage (quantum efficiency larger than 35\% and photocathode coverage larger than 75\%). \par
The sphere is enclosed by a water pool (44 m deep and 43.5 m high) that will be used as Cherenkov veto and as shield for environment radiation. On top of the water pool, there is another muon detector to accurately measure the muon tracks, described in subsection \ref{sub:veto}.

\subsection{The JUNO PMT systems}

The JUNO detector have two system to read the light from the scintillator (the so called Large PMT (LPMT) and the so called Small PMT (SPMT)), that are complementary. The LPMT system is formed by 18000 20$"$ diameter PMTs, whereas the SPMT is formed by 25000 3$"$ diameter PMTs. The two PMT systems have a coverage larger than 78\%. The LPMT will have a large dynamical range from one photo electron (p.e.) up to 100 p.e., therefore they can be affected by non-linearity effects while the SPMT will be in the counting regime (1-0 p.e.). \par
The LPMT system will provide a coverage of about 75\%, while the SPMT system a coverage of 3\%; that is why, the stochastic term will be around 3\% at 1 MeV for the LPMTs and 14\% at 1 MeV for the small ones. The SPMTs will be faster and characterized by a better single p.e. resolution than the large ones and they will have a smaller dark noise with respect to the large PMTs.

\subsection{The JUNO electronics}

\begin{figure}[htb]
\centering
\includegraphics[height=2.8in]{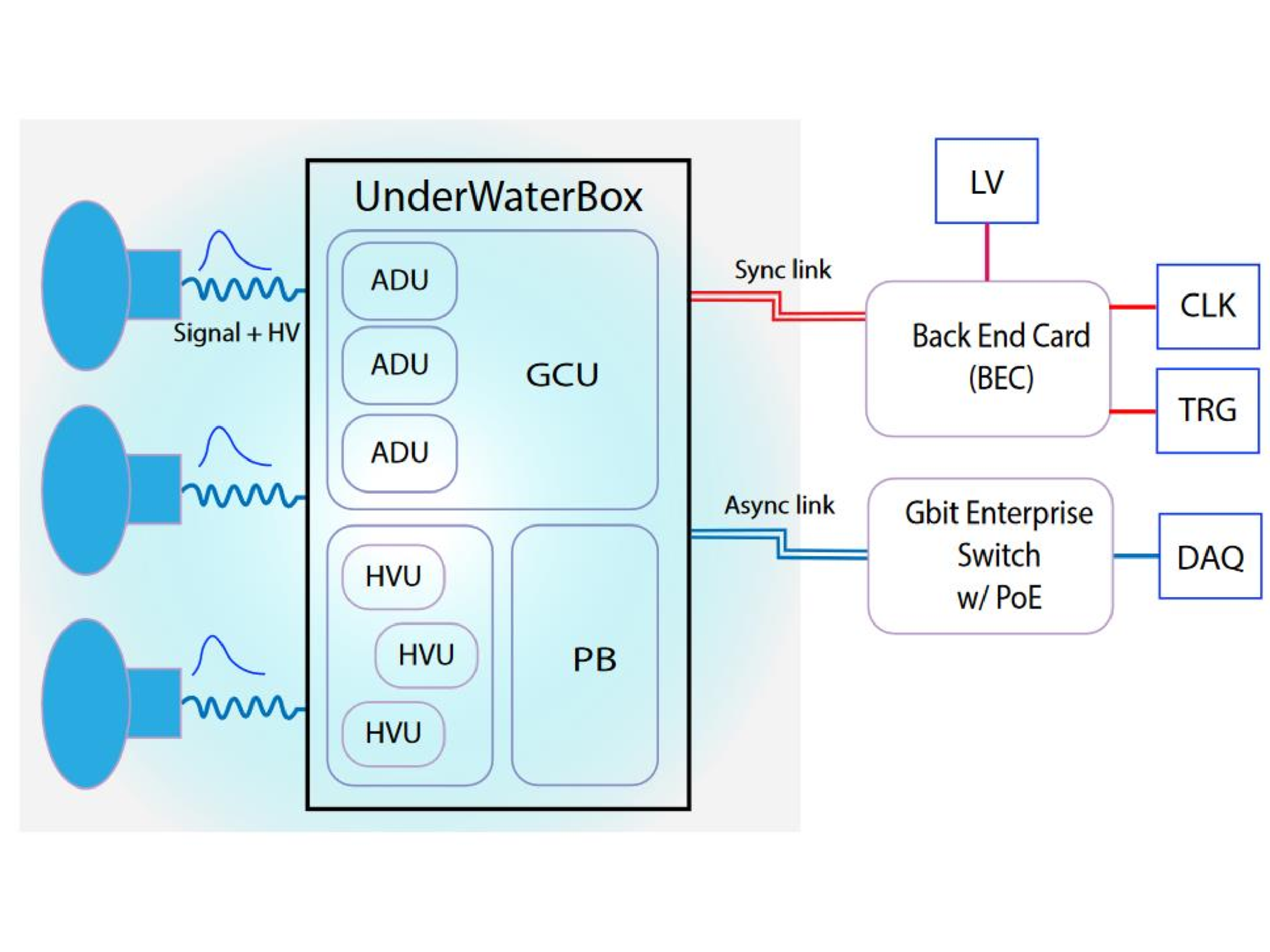}
\caption{A schematic view of the electronics of 1F3 scheme. The 3 PMTs with their bases connect with 1 m cable to the underwater box are shown on the left. The underwater box is shown in the center. It will contain the GCU plus the 3 ADU boars, a PB and the 3 HV units. On the right there is the out of water electronics connected with the 100 m Ethernet cables.}
\label{fig:1F3}
\end{figure}

The read-out electronics of the JUNO 20$"$ PMT will be inaccessible after the installation. Thus, it is designed to minimize the number of channels, the dissipated power, the number of cables and waterproof connectors and to maximize its reliability. To fulfill all these requirements the scheme of the electronics was changed on July 2017 passing from BX scheme (1 electronics channel for each PMT potted at the PMT base) to 1F3 scheme (1 electronics channel for 3 PMTs connected with 1 m cable from the PMT base to an underwater box). I would just describe the new scheme. I would not mention the BX scheme, but just describe the 1F3. In the 1F3 scheme only the base will be integrated in the PMT housing and it will be connected 1 m cable with an underwater box in which there will be the GCU (global control unit) with on it the 3 ADU units and integrated with the PB (power board) and the 3 HV units. In both schemes the signal is driven out of the water by $\sim$ 100 m of Ethernet cables synchronous link to a BEC (back and card). In the 1F3 scheme a second Ethernet cable will be connected to a Gbit enterprise switch to connect the electronics with the DAQ. A schematic view of the 1F3 scheme is shown in Fig. \ref{fig:1F3}. The front-end underwater electronics does not include only the analogue to digital conversion (ADC) stage but also intelligence of the electronics. A lot of functionalities will take place such as selective data readout, HV control, system monitoring, baseline control, signal conditioning and digitization, trigger primitive generation, segment buffering, calibration and synchronization tasks.

\subsection{The JUNO calibration systems}

The challenge of the JUNO calibration system is to control the energy scale uncertainty less than 1\% to reach the required energy resolution of $3\%/\sqrt{E(MeV)}$, therefore the calibrazion system is a crucial issue. There are four complementary methods under developing, which can perform 1D, 2D and 3D calibration for JUNO.
The four calibration methods are as follows: i) ACU: Automatic Calibration Unit, one dimensional for the central axis scan; ii) CLS: Cable Loop System, two dimensional which will scan the vertical planes inside of the acrylic sphere; iii) GTCS: Guide Tube Calibration System, two dimensional for the acrylic outer surface scan; iv) ROV: Remotely Operated under-LS Vehicle, three dimensional for the whole detector scan. \par
Different radioactive sources will be used during the calibration, in particular gamma ray (e.g. $^{40}K$, $^{54}Mn$, $^{60}Co$, $^{137}Cs$), positron (e.g. $^{22}Na$, $^{68}Ge$) and neutron sources (e.g. $^{241}Am-Be$, $^{241}Am-^{13}C$ or $^{241}Pu-^{13}C$, $^{252}Cf$).

\subsection{The veto detector} \label{sub:veto}

A veto detector is needed to: i) reduce the cosmogenic isotope of $^{9}Li/^{8}He$ by precision reconstruction of muon track; ii) reject fast neutron background by passive shielding and possible tagging; iii) shield the radioactivity from rock by water.
So the veto detector need a water Cherenkov detector and a muon top tracker installed on the top of the water pool to accurately measure the muon direction. The water Cherenkov detector is the pool in which the sphere of liquid scintillator is inserted, while the top tracker will re-use the OPERA's Target Tracker which is made of plastic scintillators.


\section{Physics with JUNO}

The main goal of the physics program with JUNO detector is the determination of the neutrino mass hierarchy with good sensitivity (3 $\sigma$ after 6 years) and the precise measurement the neutrino mixing parameters. Indeed, the $sin^2\theta _{12}$ parameter is current know with a precision 4.1\%, $\Delta m_{21}^2$ with 2.3\%, and $\Delta m_{ee}^2$ with 1.6 \%. With the JUNO experiment we expect to measure all these parameters with a precision below 1\%. \par
Many other neutrino physics aspects can be studied with JUNO detector such as supernova burst and diffuse supernova neutrinos that will provide information on standard neutrino model. A Specialized trigger under study to acquire the supernova events. From solar neutrinos it is possible to get information about on $^{7}Be$ flux and $^{8}B$ one; while from geo-neutrinos the information on U/Th ratio can be obtained, in particular a precision of 17\% on U+Th flux within the first year and of 6\% after 10 years is foreseen. Moreover, JUNO will provide also information on the indirect dark matter searches and it will exploit physics of atmospheric neutrinos. Furthermore by studying the nucleon decay, exploiting the water Cherenkov detector it will be possible to have information on proton decay $p \rightarrow K^+ + \bar{\nu}_e $.

\section{Summary and schedule}

JUNO is a multipurpose reactor neutrino experiment under construction in the south of China. The founding of JUNO project was approved in 2013 and the JUNO collaboration was officially formed in 2014. Now the JUNO collaboration is composed by more than 70 institutes and more than 550 people. The civil construction, started on 2015, is going on, in particular the tunnels are already completed. The detector component, PMT production and testing and electronics development are in progress. The detector installation is foreseen to start in 2019 while the liquid scintillator filling and detector commissioning are foreseen in 2020. \par
The physics program of JUNO experiment involves the mass hierarchy determination on $3\sigma$ after 6 years or even better. A significant improvement in the uncertainty of $sin^2\theta _{12}$, $\Delta m_{21}^2$ and $\Delta m_{ee}^2$ will also be obtained. The JUNO experiment will provide also data from supernovae, solar, atmospheric and geo-neutrinos, proton decay and it will improve the knowledge on these fields.


\begin{thebibliography}{99}


\bibitem{DayaBay}
F.P. An et al., “Measurement of electron antineutrino oscillation based on 1230 days of operation of the Daya Bay experiment”, Phys. Rev. D, {\bf 95}, 072006 (2017).

\bibitem{Reno}
S.H. Seo et al., “Spectral Measurement of the Electron Antineutrino Oscillation Amplitude and Frequency using 500 Live Days of RENO Data”, arXiv:1610.04326.

\bibitem{tre}
Y. Abe et al., “Measurement of θ13 in Double Chooz using neutron captures on hydrogen with novel background rejection techniques”, J. High Energ. Phys. {\bf 163}, (2016).

\bibitem{Adam}
T. Adam et al., “The JUNO Conceptual Design Report”, arXiv:1508.07166 (2016).



\bibitem{An}
F. An et al., “Neutrino physics with JUNO”, Journal of Phys. G (Nucl. and Part. Phys.), {\bf 43}, id. 030401 (2016).



\end{thebibliography}
\end{document}